# "Virtual IED sensor" for df rf CCP discharges

M. Bogdanova[1,2], D. Lopaev[1], T. Rakhimova[1], D. Voloshin[1], A. Zotovich[1], S. Zyryanov[1,2]

[1]Skobeltsyn Institute of Nuclear Physics, Lomonosov Moscow State University, SINP MSU, Moscow, Russia

[2]Faculty of Physics, Lomonosov Moscow State University, MSU, Moscow, Russia

## Abstract

Ion-assisted surface processes are the basis of modern plasma processing. Ion energy distribution (IED) control is critical for precise material modification, especially in atomic-level technologies such as atomic layer etching. Since this control should be done in "real time", it requires "real-time" feedback using fast process sensors. In the general case of an industrial plasma reactor, when direct IED measurement is not possible, the IED can be estimated using the concept of a "virtual IED sensor". In this paper, a similar "virtual IED sensor" is considered using an asymmetric dual-frequency (df) rf CCP discharge as an example. It is based on a fast calculation method of the IED at an rf-biased electrode. This approach uses the experimentally measured sheath voltage waveform and plasma density (or ion flux) as input data, and also includes Monte-Carlo simulation of ion motion in the sheath to take into account the effect of ion-neutral collisions. To validate this approach, experiments were carried out using various plasma diagnostics in several gases: argon and xenon as examples of plasma with atomic ions and nitrogen as an example of plasma with molecular ions. It is shown that in all cases it is possible to obtain an adequate IED estimation, close to the experimental one, in a reasonably short time (~ tens of seconds when using a modern PC). The results obtained demonstrate the possibility of using a virtual IED sensor in real plasma processing.

## 1 Introduction

Ion-assisted surface processes are the basis of modern plasma processing technologies. Most of them require precise control of various plasma parameters. For example, material modification at atomic layer scale using techniques such as atomic layer etching requires monitoring ion flux and ion energy distribution (IED) on the surface of the material being processed. It is important to note here that monitoring implies a "real-time" operating regime, i. e. during material processing. This, in turn, requires "real-time" feedback through sensors that provide information about the process and thus make the process control truly precise.

So, on the one hand, modern plasma processing requires the development of new plasma approaches and systems to ensure an appropriate level of control. On the other hand, this control requires the development of fast sensors.

The dual-frequency (df) rf ICP and CCP discharges are widely used today in plasma surface treatment. The simultaneous use of high-frequency and low-frequency rf voltages is aimed at realizing separate control of ion flux and energy. This idea was presented in the works of Goto in the early 1990s [1,2]. Since then, these discharges have been extensively studied both experimentally and theoretically [3,4]. However, one of the main results of these studies is that the two frequencies are generally coupled and the desired functional separate control can only be achieved over a limited range of discharge parameters (frequencies, pressures, and input powers).

The multi-frequency discharge excitation was also investigated [5] to reveal the role of the third frequency in controlling the IED shape [6]. In order to better control the ion energy, some new methods of rf discharge excitation have recently been developed. One of them is the excitation of a symmetric rf CCP discharge with two frequencies (the fundamental frequency and its second harmonic) with a phase shift between them. In this configuration, the dc self-bias voltage appears due to the electrical asymmetry effect (EAE) [7]. This voltage is adjusted by the phase shift value between the two frequencies and is used for additional control of ion energy. This idea was further developed in the so-called "tailored-waveform" rf discharge excitation, when the rf voltage waveform is composed of multiple rf harmonics in order to obtain the desired IED. This concept is detailed in the review paper [8]. The pulsed df discharge excitation can also provide additional control of plasma parameters, including ion energy. Some results on the ion energy control in a pulsed df ICP discharge are discussed in [9–11].

As mentioned above, in addition to choosing a suitable plasma system for processing materials, it is equally important to obtain information in "real time" about the plasma process and, preferably, directly about the flux and energy of ions on the surface. Since the direct use of invasive plasma diagnostics is not possible in industrial plasma reactors, diagnostic methods such as "virtual sensors" are being investigated. The concept of a "virtual sensor" is a combination of "real-time" experimental measurements, usually of external plasma parameters, with fast numerical simulations using these measurements as input or reference data. It works like a fast plasma model that tracks the state of the plasma, or rather the state of selected parameters, such as, for example, flux and energy of the ions at the electrode.

Experimental studies combined with numerical simulations provide better diagnostics for rf discharges. Such approach was used, for example, in our previous papers to determine the plasma density from the ion current to the Langmuir probe: using the PIC MCC simulation for the cylindrical probe [12] and the flat probe [13], and with simplified numerical approaches for these probes [14]. The relation between ion flux and plasma density in a df rf CCP discharge in various gases has been investigated both experimentally and using the 2D PIC MCC simulation in [15] to obtain the pressure dependence of this relation. The ion composition and energy spectra were studied experimentally and numerically in Ar/$H_2$ plasma of a df rf CCP discharge in [16].

All these studies were aimed at establishing relations between different plasma parameters and discharge conditions. These relations are of value for ion diagnostics, since the rf-bias voltage waveform at the electrode, although important, does not completely determine the ion energy spectrum. The ion transit time through the plasma sheath relative to the rf period and the time between ion-neutral collisions is the most important parameter that strongly influences the IED shape. The ion transit time depends on the plasma sheath thickness, which in turn is determined by the plasma density and the applied rf-bias. The time between collisions is determined by pressure. In the case of a wide collisionless plasma sheath (low pressure and low plasma density), simple analytical equations can be applied to obtain a relation between the external parameters of the rf discharge and the IED shape [5,17–26].

Some previously developed methods for fast IED calculation [5,26] only take into account the collisionless ion motion. In [5], the analytical model uses a linear transfer function to relate the time-varying sheath voltage to the time-varying ion energy response at the electrode surface. In [26], a similar approach is extended by an equivalent circuit representation of the system and an equation for the "damped" sheath potential, to which the ions respond. In these papers, ion-neutral collisions in the sheath are neglected. Therefore, such models [5,26] are applicable only at low pressures, when the ion mean free path is greater than the sheath thickness.

These fast IED calculation methods may well be taken as the basis for a virtual IED sensor in the appropriate conditions of their applicability. Recently, a similar approach has been validated for the low-pressure df rf ICP discharge in [17]. However, in the general case of intermediate-pressure plasma, when there are ion collisions in the sheath, the fast IED calculation is a more complicated problem.

In this paper, the concept of a "virtual IED sensor" is considered using the example of an asymmetric df rf CCP discharge. It should be noted that the IED, i. e. the ion energy distribution, in this work is understood as the energy distribution of the ion flux, which is directly measured by analyzers [16]. The virtual IED sensor is based on the fast calculation method of the IED at an rf-biased electrode. This approach uses the experimentally measured sheath voltage waveform and plasma density (or ion flux) as input data, and also includes Monte-Carlo simulation of ion motion in the sheath to take into account the effect of ion-neutral collisions when estimating the IED. To validate this approach, experiments were carried out using various plasma diagnostics in several gases: argon and xenon as examples of plasma containing atomic ions; and nitrogen as an example of a plasma with molecular ions.

This paper is organized as follows. The experimental setup and all diagnostics used are described in Section 2. A model for fast IED calculations is presented in Section 3. Section 4 is devoted to a discussion of the results. Conclusion is given in the last Section 5.

## 2 Experiment

### 2.1 Setup and conditions

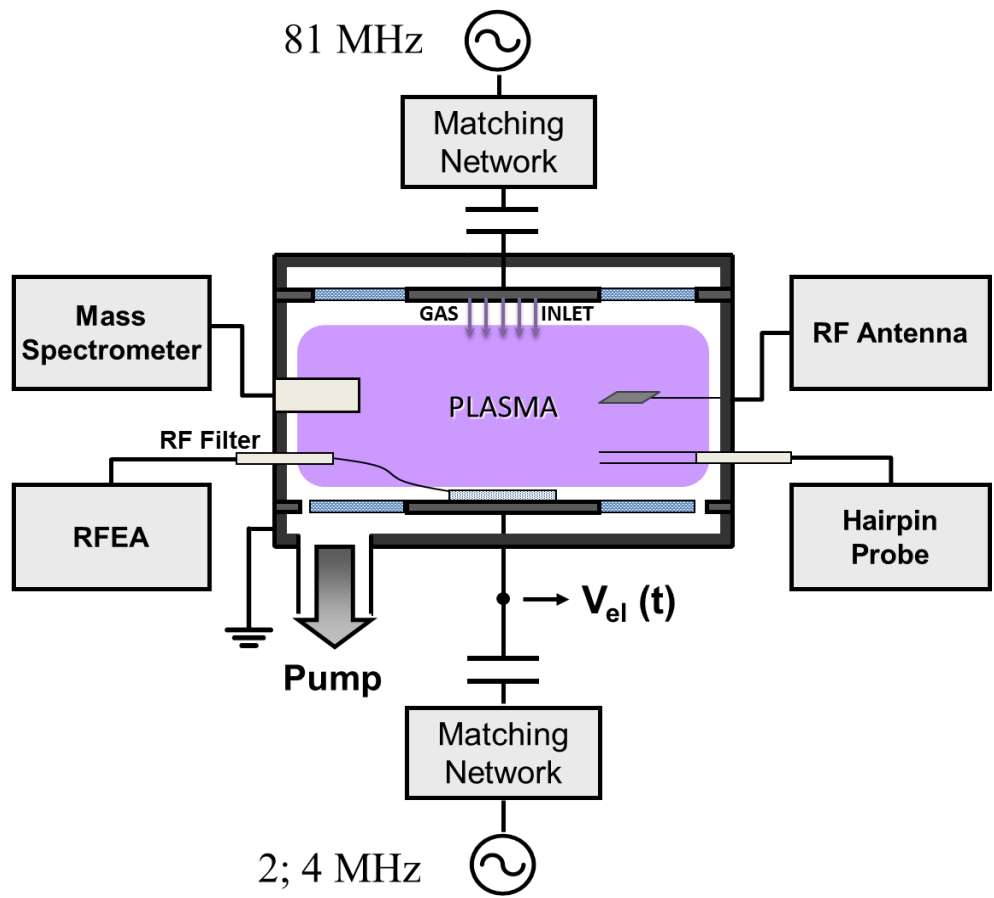

**Figure 1.** Experimental setup.

Experiments to validate the concept of the virtual IED sensor are carried out in an asymmetric dual-frequency (df) rf CCP discharge [15,16]. It is schematically shown in figure 1 along with the diagnostics used. The discharge chamber made of stainless steel has a cylindrical form. Its sidewalls are 45 mm high, and bases are 330 mm in diameter. Al electrodes (120 mm in diameter) are placed at its bases at a distance of 30 mm from each other. Both electrodes are surrounded with the quartz rings with the external diameter of 240 mm.

The df rf CCP discharge [3,5] is used to separately control the ion flux and energy. Here, a high-frequency (81 MHz) voltage is used for plasma generation. A voltage of 81 MHz is applied to the top electrode through a blocking capacitor. An input power of 81 MHz determines the plasma density $n_e$ and ion flux $F_i$ to the bottom electrode. The bottom electrode and sidewalls are grounded at a frequency of 81 MHz ($Z_{81} \to 0$). Since its total surface area exceeds the surface area of the top electrode, the CCP discharge at 81 MHz is geometrically asymmetric. It is essential that in this case the plasma potential $V_p$ hardly changes even with a significant change in the input power at 81 MHz.

A low-frequency (2 and 4 MHz) rf-bias voltage $V_{rf-bias}(t)$ is applied to the bottom electrode to adjust the ion energy distribution (IED). As the top electrode and sidewalls are grounded at low frequencies, by analogy with the high-frequency case, the discharge is asymmetric. The rf-bias voltage parameters – frequency $f_{rf-bias}$ and amplitude $V_{rf-bias}$ – strongly affect the sheath above the bottom electrode, but do not affect plasma bulk. Ions, moving to the rf-biased electrode, gain energy in the sheath [17,20,22,23,27–30]. Thus, the IED is mostly determined by the rf-bias voltage: its frequency and amplitude as well as the waveform. To verify the virtual IED sensor concept, depending on various discharge conditions, the rf-bias parameters in the experiments are varied: frequency $f_{rf-bias}$ - 2 and 4 MHz, amplitude $V_{rf-bias}$ – 0, 30, 60 and 120 V.

To validate the concept, depending on the gas sort, Ar-, N$_2$- and Xe-plasma are investigated. Gas is injected into the discharge chamber through the gas shower in the top electrode (orifices 200 μm in diameter with the 8 mm step). The experiments are carried out at a pressure $p$ of 20 and 100 mTorr to study the virtual IED sensor performance in the slightly collisional and collisional regimes.

### 2.2 Diagnostics

Validation of the virtual IED sensor is carried out via the rf-compensated Retarded Field Energy Analyzer (RFEA), placed at the rf-biased electrode [22,23,31–34]. The IEDs, obtained by the virtual sensor, are compared with the results of the direct RFEA measurements. The fast IED calculation method [16] that underlies the proposed virtual IED sensor (detailed description may be found in Section 3) requires the following input parameters: ion composition, sheath thickness $s_m$ and sheath voltage waveform $V_s(t)$. This section provides a brief description of diagnostic methods, which are used to obtain data for input parameters and validation of the IED sensor.

*RFEA.* IEDs are directly measured by the rf-compensated RFEA, placed on the surface of the rf-biased electrode. The analyzer is a stack of three Ni grids (grid cell dimensions ~17x17 μm with an open window ~10x10 μm) and a collector plate, enclosed in an Al case (~5 mm thick and 120 mm in diameter). In the front Al plate (0.8 mm thick) on a functional area with a diameter of ~10 mm there are several tens of the entrance orifices (with a diameter of 0.8 mm). These orifices are made with a step of 1 mm, providing the ~50 % plate transparency over the functional area.

The first grid is in direct contact with the RFEA front plate. The second grid is used to analyze ion energies. The third grid and the collector plate are negatively biased (–56 V and –20 V, respectively) relative to the dc self-bias potential $-V_{dc}$ at the electrode. This is done to exclude electrons coming from the plasma as well as the secondary electrons from the collector.

The distance between the grids is ~0.3 mm. In the case of Ar-plasma, this distance provides correct operation of the RFEA at pressures up to ~200 mTorr. Rf compensation of the RFEA is provided both by sets of resonant rf filters and rf chokes, connected to the electrical circuits of the collector and grids, and by a large area of grid-plates (and accordingly, a large capacitance between grids $C_{gr} \sim 150$ pF). The electronic unit of the analyzer has a fast feedback loop for dc self-bias $-V_{dc}$, resulting in higher reproducibility and accuracy of IED measurements.

*Rf antenna.* The sheath voltage waveform $V_s(t)$ strongly affects the IED shape [7,8,28,35–40], so it is extremely important to take it into account. The fast IED calculation method requires the sheath voltage waveform $V_s(t)$ as an input parameter. Experimentally, it is obtained as the difference between the measured potentials of the electrode $V_{el}(t)$ and the plasma $V_p(t)$:

$$V_s(t) = V_{el}(t) - V_p(t). \qquad (2.1)$$

The electrode potential waveform $V_{el}(t)$ was measured by a high-voltage probe at the point marked in figure 1. The plasma potential waveform $V_p(t)$ was measured by an rf antenna ($A_{ant} \sim 6$ cm$^2$), inserted in the plasma bulk (see figure 1) and connected with a high-impedance rf probe ( $C_{probe} \sim 1.5$ pF, $f_{cut} \sim 100$ MHz).

Typical electrode $V_{el}(t)$ and plasma $V_p(t)$ potential waveforms, measured in Ar-plasma at a pressure of 20 mTorr and at rf-bias parameters of 2 MHz and 60 V, are demonstrated in figure 2(a). The plasma potential waveform $V_p(t)$ indicates that the amplitude of low-frequency oscillations in plasma $V_p^{2\ MHz}$ is much less than the amplitude of high-frequency oscillations $V_p^{81\ MHz}$. It is also much less than the amplitude of low-frequency oscillations at the electrode $V_{el}^{2\ MHz}$, to which the low-frequency rf-bias is applied. Based on this, it can be concluded that there is a frequency decoupling [1–3,8,41,42] in the plasma bulk, i. e. the applied rf-bias at a low frequency has almost no effect on the plasma bulk parameters.

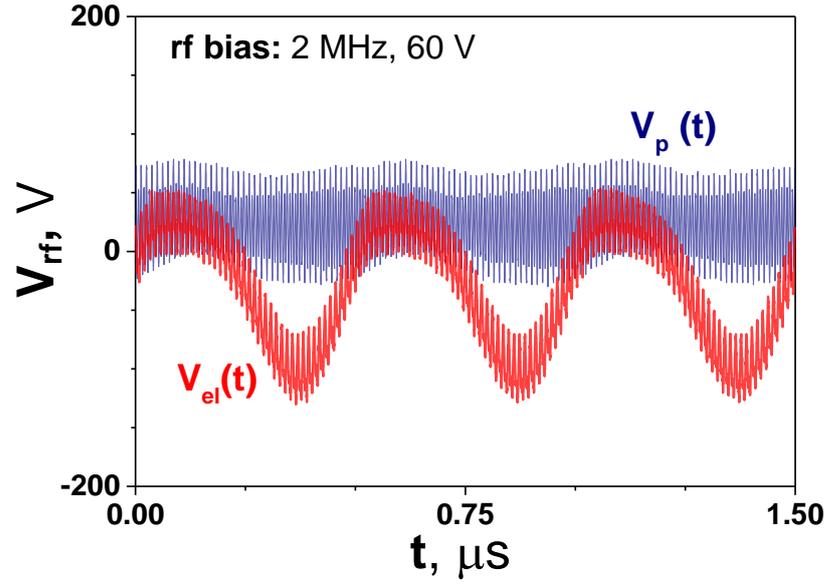

(a)

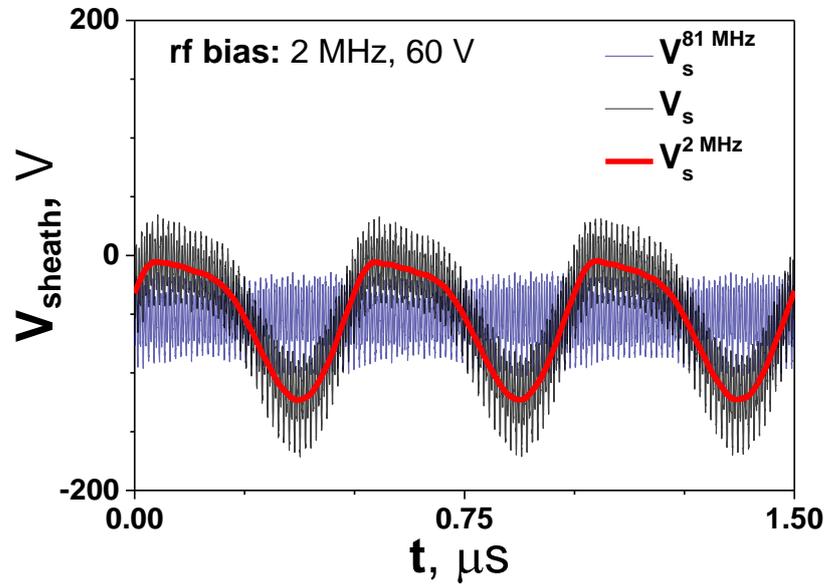

(b)

**Figure 2.** Typical electrode $V_{el}(t)$ and plasma $V_p(t)$ potential waveforms measured at an applied rf-bias of 2 MHz and 60 V in Ar-plasma at a pressure of 20 mTorr – (a) and the corresponding sheath voltage waveform $V_s(t)$ with its high- and low-frequency components - $V_s^{81\,MHz}(t)$ and $V_s^{2\,MHz}(t)$, respectively – (b).

The corresponding sheath voltage waveform $V_s(t)$ is shown in figure 2(b) along with its low-frequency $V_s^{2\,MHz}(t)$ and high-frequency $V_s^{81\,MHz}(t)$ components. It is worth noting here that the low-frequency $V_s^{2\,MHz}(t)$ component is not sinusoidal.

*Hairpin-probe.* The experiments are carried out in plasma of constant plasma density $n_e$ (Ar- and Xe-plasma density is $2 \cdot 10^{10}$ cm$^{-3}$, N$_2$-plasma density is $3 \cdot 10^{9}$ cm$^{-3}$). Thus, the sheath thickness $s_m$ in plasma of each sort is determined only by the sheath voltage $V_s$ at a certain pressure.

As there is a frequency decoupling in the plasma bulk, the plasma density $n_e$ is determined by the high-frequency power input at 81 MHz. The $n_e$ value is kept constant as follows: the current $n_e$ value is monitored by the hairpin-probe [43–45], and when changing the gas sort or pressure, it is adjusted by varying the input power at 81 MHz.

A hairpin-probe, made of a tungsten wire (~ 10 cm long, 200 μm in diameter), is introduced into the plasma bulk through a sidewall diagnostic window. The distance between the hairpin tips is 8 mm. The cable with loop antenna, which excites the oscillations in the hairpin, is placed inside a thick glass tube to minimize its influence on plasma.

*Mass-spectrometer.* The ion composition of plasma is monitored using the Hiden Analytical EQP 300 mass-spectrometer. Its diagnostic head with the entrance orifice (50 μm in diameter) is located between the electrode edge and the sidewall at a height of 15 mm above the bottom edge of the electrode (which corresponds to half the interelectrode distance).

The ion composition measurements have revealed the insignificant admixture of $H_2O^+$: < 6% in Ar, < 2% in $N_2$ and < 1% in Xe. The higher percentage of water ions in Ar- and $N_2$-plasma is caused by the fast charge exchange reaction of $Ar^+$ and $N_2^+$ ions with $H_2O$ molecules, which have a lower ionization potential. Thus, a significant part of the water is in the ionic state. For Xe, the ionization potential is very close to the potential for $H_2O$, and therefore the percentage of water ions in Xe turns out to be very small. The $H_2O^+$ admixture is not taken into account, i. e. it is assumed that each of the plasmas used has a simple ion composition. The main ions are: $^{40}Ar^+$ in Ar ($^{41}ArH^+$ < 15 %), $^{28}N_2^+$ in $N_2$ ($^{29}N_2^+$ < 15%). The complex isotopic composition is observed in Xe. The main isotopes, $^{129}Xe^+$, $^{131}Xe^+$, $^{132}Xe^+$, are almost equal fractions and in total are ~ 70% of the Xe-plasma ion composition, whereas $^{128}Xe^+$, $^{130}Xe^+$, $^{134}Xe^+$, $^{136}Xe^+$ are the rest ~ 30%. The main ion mass was calculated from the measured isotopic composition: $M_i = 131$ amu.

## 3 Model

The virtual IED sensor concept is based on the fast IED calculation method, described in [16]. It is a method for calculating the ion motion in the sheath using experimentally measured electric field potentials and the Monte-Carlo collision (MCC) algorithm for ion collisions with neutrals.

The experimental sheath voltage waveform $V_s(t)$ is used as an input parameter. An example of this waveform is shown in figure 2(b) along with the high-frequency $V_s^{81\ MHz}(t)$ and the low-frequency $V_s^{2\ MHz}(t)$ components.

The sheath edge motion $s(t)$ is calculated using $V_s(t)$ by the formula:

$$s(t) = s_m(1 - |V_s(t)/V_{max}|), \qquad (3.1)$$

where $V_{max}$ is the maximum sheath voltage and $s_m$ is the maximum sheath thickness. The sheath thickness during an rf period varies as $(s_m - s(t))$. The sheath edge $s(t)$ is counted from the plasma bulk, and its zero $s(t) = 0$ corresponds to $s_m$ when $V_s(t) = V_{max}$. The sheath edge is closest to the electrode when the sheath voltage is minimal.

For the electric field $E(x,t)$ in the sheath, an expression corresponding to the linear sheath model [19] is used:

$$E(x,t) = \begin{cases} \frac{2V_s(t)}{s_m^2}(x - s(t)), & x \geq s(t) \\ 0, & x < s(t) \end{cases} \qquad (3.2)$$

The maximum sheath thickness $s_m$ can be found from the formula for the collisionless Child-Langmuir sheath [19]:

$$s_m^2 = K_i \varepsilon_0 \left(\frac{2e}{M_i}\right)^{1/2} \frac{\bar{V}_s^{3/2}}{\bar{J}_i}, \qquad (3.3)$$

where $e$ is the electron charge, $\varepsilon_0$ is the dielectric constant, $K_i$ is a numerical constant equal to 4/9 for a dc sheath and approximately 0.82 for an rf sheath, i. e. in the case when electrons rf periodically fill the sheath; $\bar{J}_i$ is the average ion current density (measured experimentally), $\bar{V}_s$ is the time-average sheath voltage, $M_i$ is the ion mass. The formula (3.3) is applicable in the case of a high-frequency discharge, or a dual-frequency discharge with the so-called "current-driven regime", when the sinusoidal rf current density passes through the sheath. In a single-frequency low-frequency discharge, when there is a significant ion current modulation and, therefore, a non-sinusoidal rf current density over the rf period, formula (3.3) cannot be applied.

The question of estimating $s_m$ becomes more complicated if the sheath voltage waveform is not sinusoidal and collisions occur in the sheath, which is much more often observed in real plasma and therefore is considered in this work. Since this parameter determines the ion transit time through the collisionless sheath, and if it is collisional, then the number of collisions in it, it is necessary to choose some $s_m$ value for the numerical model to give a correct estimate of the IED.

In some papers the sheath thickness $s_m$ value is obtained by its variation until the calculated IED becomes similar to the experimentally measured one [46]. This approach is illustrative for studying the IED dependence on $s_m$. However, the purpose of this work is to validate a method for calculating IED that does not require direct measurements of the IED. But we will use the variation of the $s_m$ value to find the best fit between the calculated IED and the measured directly, and then compare the obtained $s_m$ results with those, calculated by formula (3.3) with different $K_i$. The results of this comparison will reveal the most appropriate way to estimate $s_m$ under the considered discharge conditions.

The ion trajectories in a given electric field are calculated by the Monte-Carlo collision method. The motion of $10^5$ particles injected into the sheath at different phases of the rf period is traced. This number of particles provides an IED calculation with good statistics in a relatively short time. A further increase in the number of particles does not change the IED shape, but increases the computation time. The influence of different statistics ($10^5$, $10^4$ and $10^3$ particles) on the accuracy and time of IED calculation is discussed in Section 4.

Collisions are implemented using a standard null-collision approach [47]. The simulation uses the experimental voltages measured at rf-bias parameters of 2 and 4 MHz and 0, 30, 60 and 120 V. Ar and $N_2$ are simulated at two pressures – 20 and 100 mTorr, while for the Xe case only 20 mTorr is used.

The isotropic part of elastic collisions and backward scattering part (charge exchange) of ion elastic collisions with atoms are included in the model with the cross-sections for $Ar^+$ from [48], for $N_2^+$ from [49] and for $Xe^+$ from [50].

The simulation should use the sheath voltage $V_s(t)$ measurements over several rf-bias periods. Despite the $V_s(t)$ periodicity, the IED calculated using multiple rf-bias periods of $V_s(t)$ is much smoother than the IED calculated using a single $V_s(t)$ rf-bias period. In our simulations, we use 4 periods for 2 MHz and 8 periods for 4 MHz.

The reason for this is that the measured sheath voltage waveform $V_s(t)$ differs slightly from one period to another. Possible errors in measurements of high-frequency voltage components can also be reduced by using more rf periods.

The actual time required to calculate one IED with the proposed model ranges from one to several minutes for $10^5$ ions in a single-threaded mode on a relatively modern CPU. To use this model as a "real-time" virtual IED sensor during the experiment, $10^4$ ions can be used, then the computation time of the corresponding IED will be reduced by 10 times. The possibility of the "real-time" operating regime is discussed in Section 4. Despite the use of the Monte-Carlo method, it can be considered as a fast model compared to analogous IED simulations with the self-consistent electric field calculations (few hours/days for an rf CCP discharge simulation).

## 4 Results & Discussion

The accuracy of the virtual IED sensor is proposed to be determined by how well the sensor can reproduce the IED shape depending on the gas sort, pressure and rf-bias parameters.

The IEDs obtained by the sensor in Ar-, N$_2$- and Xe-plasma are compared with those directly measured with an RFEA, placed on the rf-biased electrode.

Two pressures are considered – 20 and 100 mTorr. The lower pressure corresponds to the slightly collisional sheath, where the IED shape is determined primarily by the ion response to the rf-bias, and the impact of collisions on it is negligible. This response is determined by the sheath voltage waveform $V_s(t)$ [16], which may differ markedly from the sinusoidal one (see figure 2(b)). At higher pressures, the effect of collisions on the IED shape is much more pronounced and is determined by the gas sort. Virtual IED sensor and RFEA data at 20 mTorr are used to test the sensor response to rf-bias parameters, and the 100 mTorr regime is used to reveal its sensitivity to collisions in the sheath.

One of the key plasma parameters influencing the IED shape is the sheath thickness $s_m$, which cannot be measured directly. Estimating its value in the case of a discharge with an asymmetric geometry and an applied low-frequency rf-bias is a nontrivial task. The influence of the $s_m$ estimate on the virtual IED sensor accuracy is discussed below.

The sheath voltage waveform $V_s(t)$ is used by the virtual IED sensor as an input parameter. In this work it is measured as the difference between the electrode $V_{el}(t)$ and plasma $V_p(t)$ potentials (2.1). The former may be easily measured with an rf-probe, while the latter requires an rf-antenna inserted into the plasma bulk (see section 2.2). It is quite obvious that the use of this rf-antenna is inconvenient, since the fewer diagnostics are placed inside the discharge chamber, the better for plasma processing. Some considerations are given as to how important it is to measure the potential waveforms of both the electrode $V_{el}(t)$ and the plasma $V_p(t)$ for proper sensor operation and how the $V_s(t)$ data accuracy affects the accuracy of the IED estimation.

Finally, the speed of the sensor, i. e. the time it takes to estimate one IED, is important when implementing the "real-time" operating regime. Therefore, the challenge is to find a balance between the accuracy of the IED estimation and the time it takes to get it. Obviously, one of these sensor characteristics will have to be sacrificed in favor of another.

Some clarifications on the mentioned points are presented below.

**4.1 Slightly collisional sheath**

*Ar.* The IEDs, obtained with the virtual IED sensor and measured with the RFEA in Ar-plasma, are presented in figure 3. The data, corresponding to rf-bias frequencies of 4 and 2 MHz, are shown in figures 3(a) and (b), respectively.

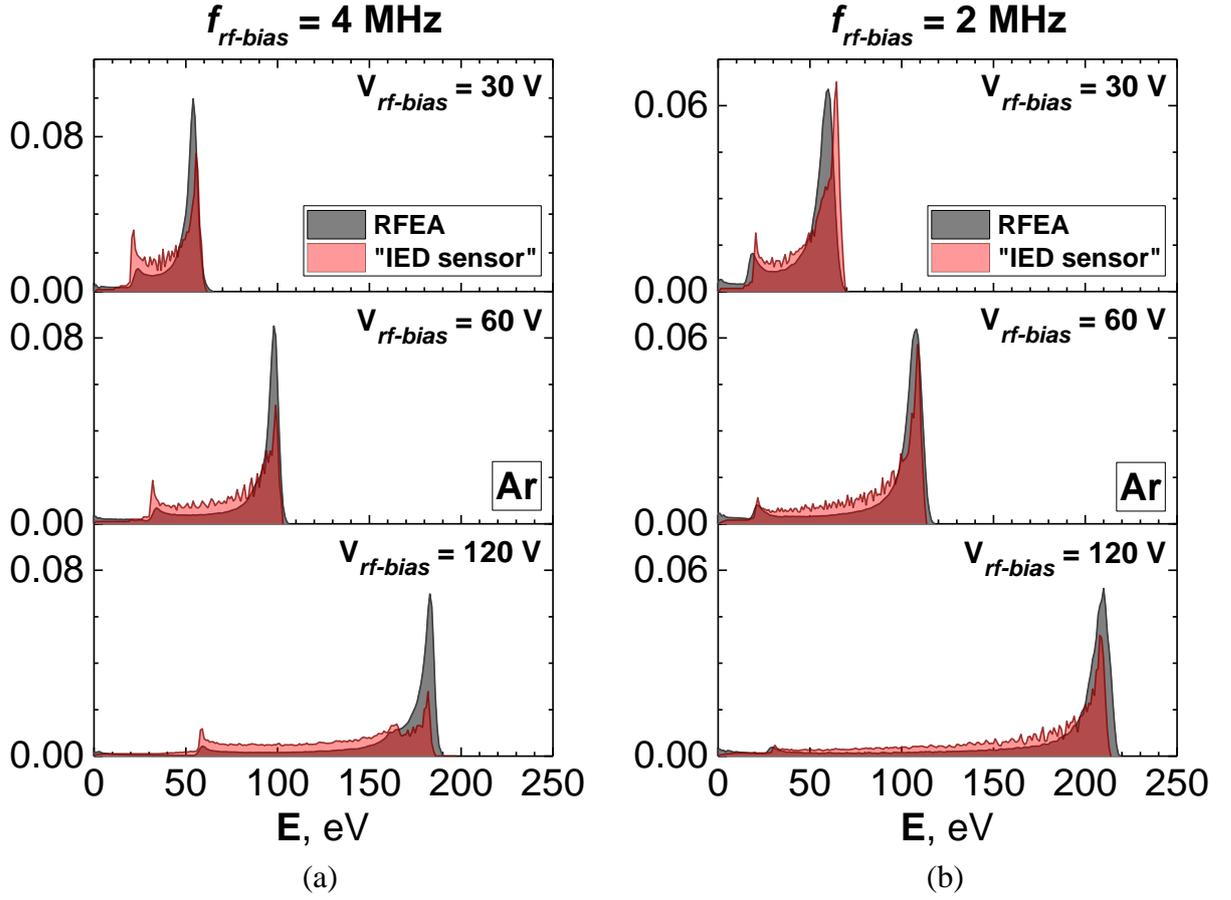

**Figure 3.** The virtual IED sensor validation in Ar-plasma ($n_e = 2 \cdot 10^{10}$ cm$^{-3}$) at a pressure of 20 mTorr and different rf-bias parameters: (a) $f_{rf-bias} = 4$ MHz; (b) $f_{rf-bias} = 2$ MHz. The RFEA measurements are grey, the virtual IED sensor data are red.

The virtual IED sensor reproduces well the IED shape and the main trends depending on the rf-bias parameters [17,18,20,22,23,27–30], namely:

- the lower the rf-bias frequency $f_{rf-bias}$ (when the rf-bias voltage amplitude $V_{rf-bias}$ is fixed), the wider the IED (spectra in figure 3 from left to right);
- the higher the rf-bias voltage amplitude $V_{rf-bias}$ (when the rf-bias frequency $f_{rf-bias}$ is fixed), the wider the IED (spectra in figure 3 from top to bottom).

Figure 3 also shows that the "asymmetry" of the IED shape (i. e. the peak amplitude inequality of the "simplest" bi-modal IED shape [16]) is well reproduced by the sensor due to the measured sheath voltage waveform $V_s(t)$ (see figure 2) used as an input parameter.

*N$_2$.* The results for N$_2$-plasma at 20 mTorr are shown in figure 4 in the same way as for Ar-plasma. It can be seen that the sensor reproduces well the IED trends depending on the rf-bias parameters. However, there are several differences in the IED shapes in Ar- and N$_2$-plasmas. In the case of N$_2$-plasma:

- the IED shape has less pronounced asymmetry of the main peaks;

- almost all IEDs show the presence of small secondary peaks.

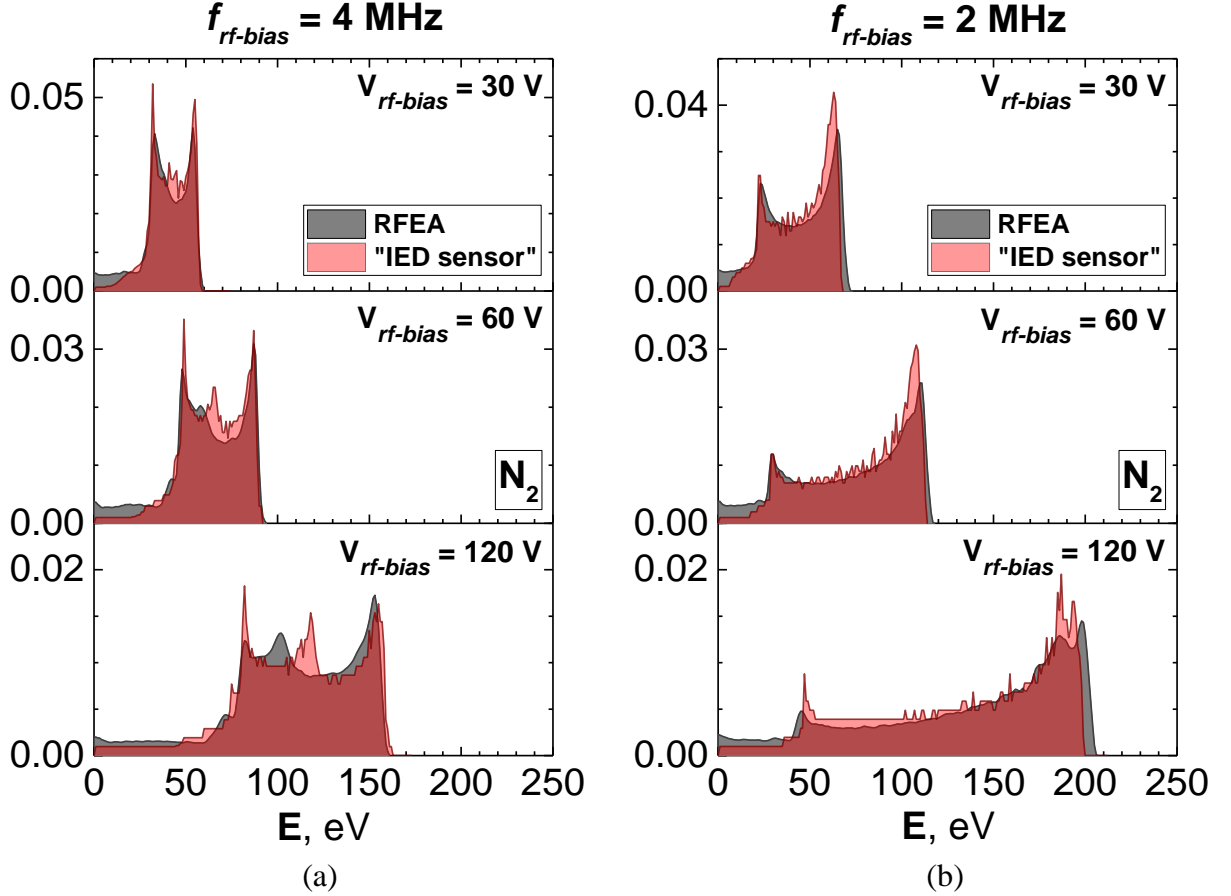

**Figure 4.** The virtual IED sensor validation in N$_2$-plasma ($n_e = 3 \cdot 10^9$ cm$^{-3}$) at a pressure of 20 mTorr and different rf-bias parameters: (a) $f_{rf-bias}$ = 4 MHz; (b) $f_{rf-bias}$ = 2 MHz. The RFEA measurements are grey, the virtual IED sensor data are red.

The first difference may be due to the fact, that the Ar-plasma density is almost an order of magnitude higher than the N$_2$-plasma density ($n_e(\text{Ar}) = 2 \cdot 10^{10}$ cm$^{-3}$, $n_e(\text{N}_2) = 3 \cdot 10^9$ cm$^{-3}$). So, the sheath thickness $s_m$ in the first case is expected to be ~ 2.5 times less than in the second one (see below in figures 9(a) and 10(a)), which, in turn, can lead to a smaller deviation of the sheath voltage waveform $V_s(t)$ from the sine.

The second difference – the presence of secondary peaks (more clearly seen in figure 4(a)) – is explained by charge exchange collisions. It is worth noting that due to the fast IED calculation method underlying the virtual IED sensor, these collisions are taken into account. The sensor reproduces these peaks, though with a slight shift on the energy scale. Despite the fact that the ion charge exchange cross-section for N$_2$ is lower than for Ar, the secondary peaks are more clearly visible in N$_2$-plasma. The reason also lies in the difference in sheath thickness $s_m$: the ions in N$_2$-plasma pass a longer way and therefore have a higher collision probability.

*Xe*. The verification of the virtual IED sensor in Xe-plasma is demonstrated in figure 5. Again, it can be seen that the sensor reproduces well all IED trends depending on the rf-bias parameters, as well as the asymmetry of the IED shape and the presence of some secondary peaks.

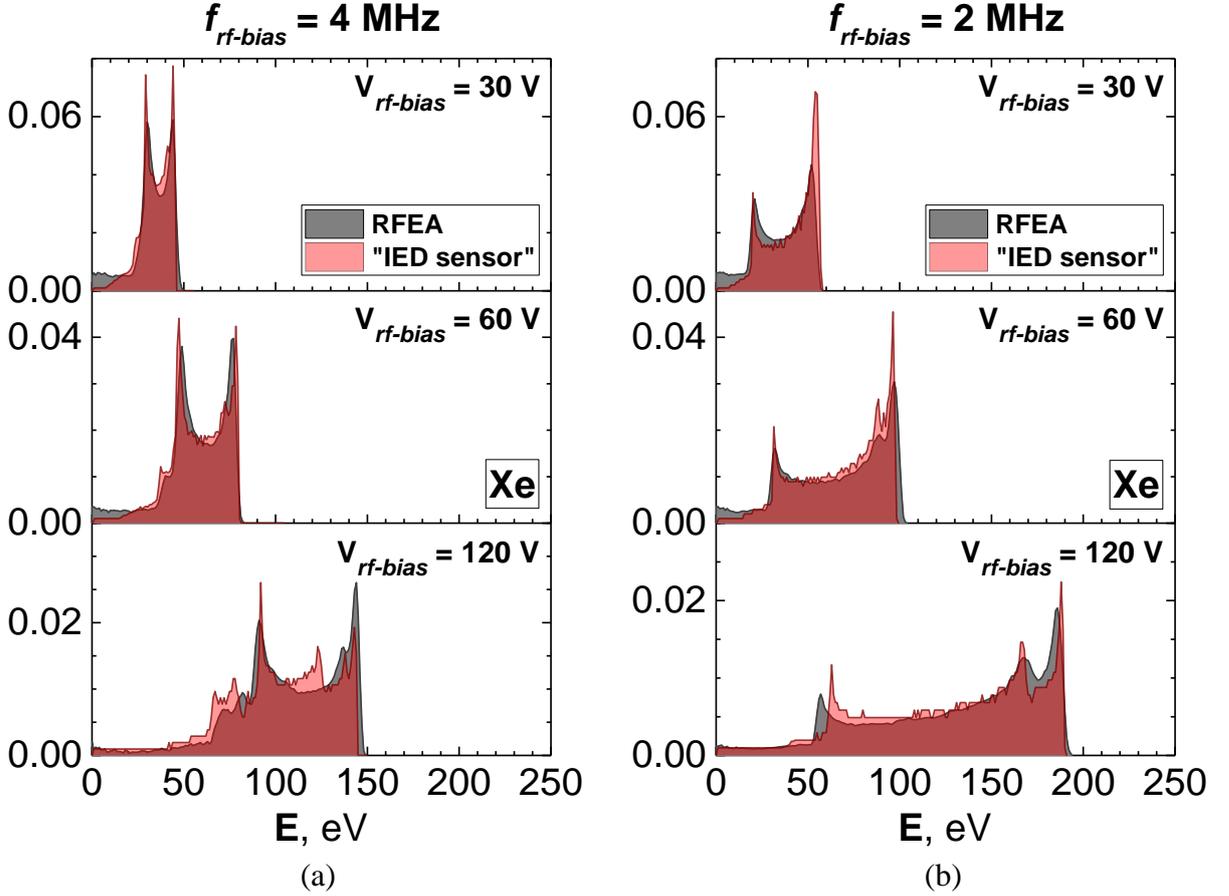

**Figure 5.** The virtual IED sensor validation in Xe-plasma ($n_e = 2 \cdot 10^{10}$ cm$^{-3}$) at a pressure of 20 mTorr and different rf-bias parameters: (a) $f_{rf-bias}$ = 4 MHz; (b) $f_{rf-bias}$ = 2 MHz. The RFEA measurements are grey, the virtual IED sensor data are red.

The shape of the IEDs is even more symmetric than in the case of N$_2$-plasma. It is assumed that the difference in the IED shapes of Ar$^+$ and N$_2^+$ arises from the difference in the sheath thickness in these plasmas. In Xe-plasma, the sheath thickness $s_m$ is close to that in Ar-plasma (see below in figures 9(a) and 11(a)). The plasma density in Ar and Xe is the same ($n_e = 2 \cdot 10^{10}$ cm$^{-3}$), but the high-frequency power input $P^{81\,MHz}$, required to obtain a plasma of this density, is much lower in the case of Xe ($P_{Ar}^{81\,MHz} \sim$ 20 W, $P_{Xe}^{81\,MHz} \sim$ 5 W). For this reason, the Xe-plasma is concentrated between the electrodes and does not have good contact with the CCP chamber sidewalls, which affects the asymmetry of the discharge. Thus, in the case of Xe, the discharge has a geometry that is closer to symmetric, which is why the sheath voltage waveform $V_s(t)$ is closer to sinusoidal than in Ar-plasma (see figure 2(b)).

As seen from figure 5, the presence of secondary peaks in the IEDs is also characteristic of Xe-plasma. For instance, in figure 5(a) with $V_{rf-bias}$ = 120 V, three secondary peaks can be distinguished on the IED, measured with the RFEA (grey), although the sensor shows even more (red).

A small computational experiment has been performed to reveal the nature of the peaks. The fast IED calculation method has been used to estimate the IED in Xe-plasma at rf-bias parameters of 4 MHz and 120 V. The gas pressure value $p$ was set so low as to completely eliminate collisional effects; therefore, the only possible nature of the peaks would be the ion response to the applied rf-bias field. The results of this experiment are presented in figure 6. For ease of comparison, the collisionless IED (labeled "$p \to 0$ mTorr") is scaled so that the rf peaks are about the same height as the IED calculated for 20 mTorr.

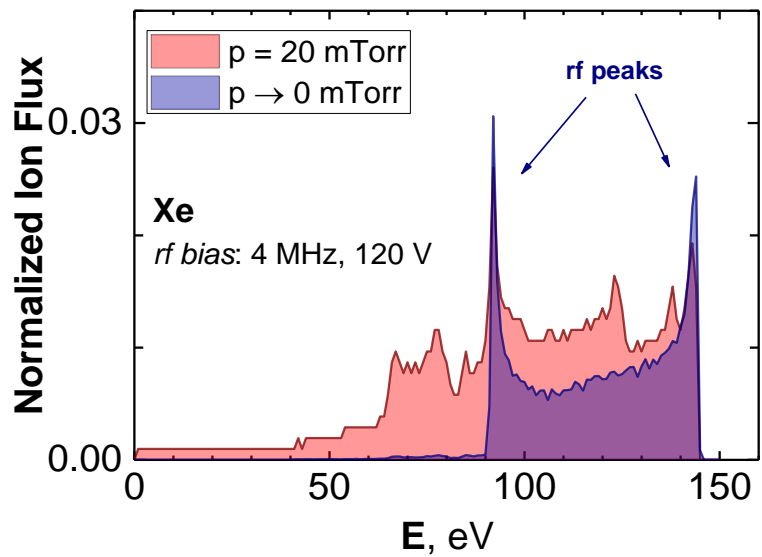

**Figure 6.** Computational experiment to reveal the nature of individual peaks in the IED. The IEDs, obtained by the fast IED calculation method for Xe-plasma ($n_e = 2 \cdot 10^{10}$ cm$^{-3}$) at rf-bias parameters of 4 MHz and 120 V at pressures of 20 mTorr (red) and close to zero (blue).

The experiment shows that the higher peaks do arise from the ion response to the rf-bias, and the rest are due to charge exchange collisions (see figure 6). It is not surprising that the Xe[+] IEDs have several charge exchange peaks even at 20 mTorr. The charge exchange cross-section for Xe is about 5 times that for Ar [48,50].

With the fast IED calculation method that takes into account ion collisions, the virtual IED sensor reproduces all peaks. This increases the sensor accuracy, which is well confirmed in the case under consideration.

## 4.2 Effect of collisions

Before discussing the virtual IED sensor performance at a higher pressure of 100 mTorr, it is essential to consider how this regime can affect the RFEA performance. The fact is that with increasing pressure, the ion mean free path $\lambda_i$ becomes shorter and can already be comparable with the distance between the RFEA grids. This means that ions can undergo collisions inside the RFEA, resulting in distorted IED measurements.

Simple estimates of the mean free path $\lambda_i$ in Ar, N$_2$ and Xe at a pressure of 100 mTorr are $\lambda_i(\text{Ar}) \sim 5 \cdot 10^{-2}$ cm, $\lambda_i(\text{N}_2) \sim 1 \cdot 10^{-1}$ cm, $\lambda_i(\text{Xe}) \sim 1 \cdot 10^{-2}$ cm. Since charge exchange collisions are predominant, estimates are carried out with the corresponding cross-sections $\sigma_{Ar} \sim 4 \cdot 10^{-15}$ cm$^2$ [48], $\sigma_{N_2} \sim 2 \cdot 10^{-15}$ cm$^2$ [49], $\sigma_{Xe} \sim 20 \cdot 10^{-15}$ cm$^2$ [50]. The mean free path of Xe$^+$ ions is $\sim 3$ times less than the distance between the RFEA grids $d = 0.03$ cm. Consequently, collisions inside the RFEA greatly distort the IEDs in Xe-plasma, making these measurements questionable. For this reason, the results of the virtual IED sensor validation at a higher pressure are presented only for Ar- and N$_2$-plasma.

It is also worth noting that the distortion of the RFEA data caused by collisions within the analyzer is difficult to account for. Apparently, this depends on the ion energy, since the lower the ion energy, the greater the number of collisions and the wider the angular distribution of the ions. Both of these factors indicate that distortion will be more pronounced in the low-energy region of the IED. And this is a reason to believe that the analyzer detects low-energy ions with a greater error [32,34].

*Ar.* The virtual IED sensor validation in Ar at a higher pressure of 100 mTorr is presented in figure 7 for the same set of rf-bias parameters. In general, the IED estimations by the virtual sensor are in good agreement with the experimental data, but there is something to discuss. In figure 7(a) the calculated IEDs have two high-energy peaks. The fast IED calculation method shows that one arises due to the rf sheath modulation and the other is due to charge exchange collisions. These two high-energy peaks are not visible in the IEDs measured with the RFEA. There is also a discrepancy in the low-energy region ($0 \div 20$ eV). This may be related to some distortion of direct IED measurements at higher pressure, which is very difficult to account for as mentioned above. The virtual IED sensor in this case appears to be better at reproducing the IED shape than the RFEA [17]. For the same reason, the number of peaks in the calculated IED may be larger and their width may be smaller compared to the IED measured by the RFEA under the same discharge conditions.

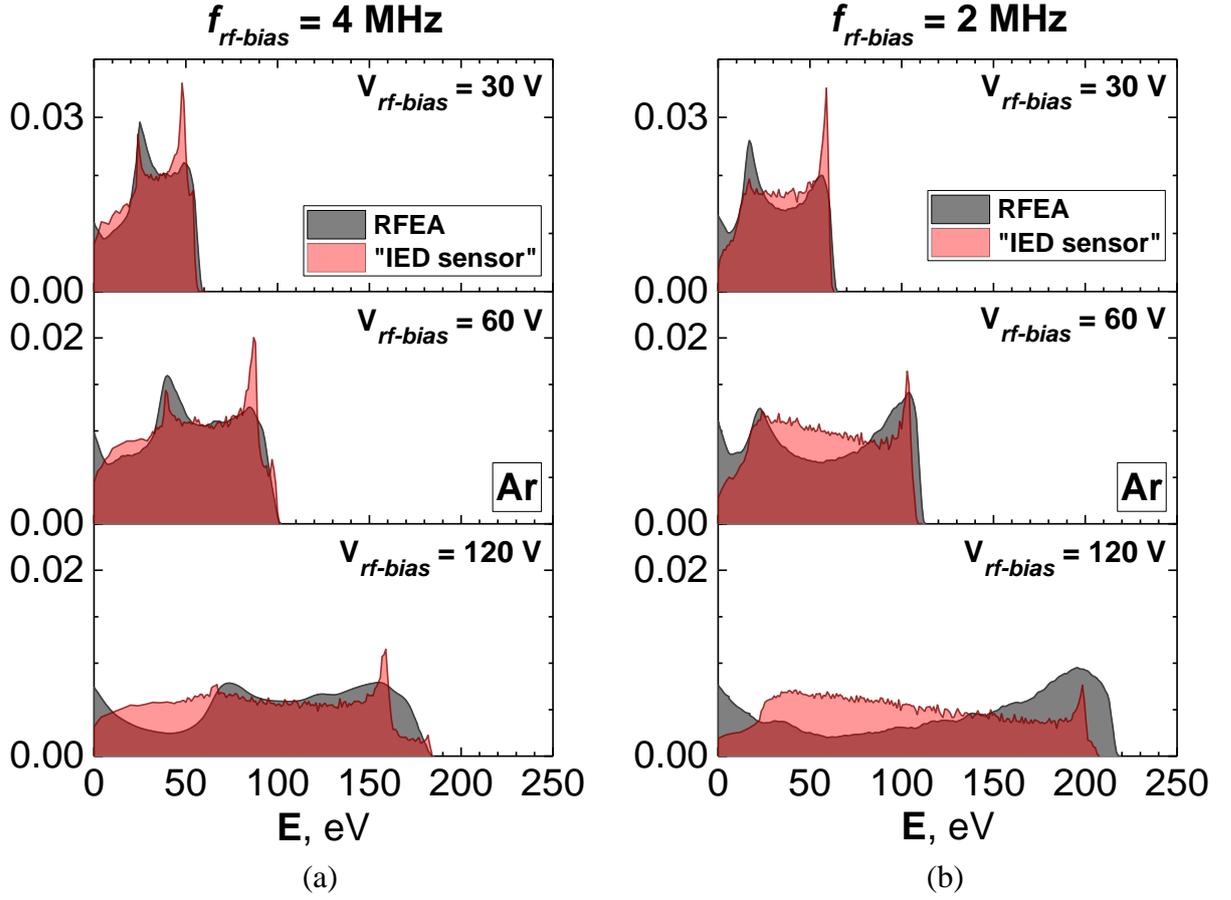

**Figure 7.** The virtual IED sensor validation in Ar-plasma ($n_e = 2 \cdot 10^{10}$ cm$^{-3}$) at a pressure of 100 mTorr and different rf-bias parameters: (a) $f_{rf-bias} = 4$ MHz; (b) $f_{rf-bias} = 2$ MHz. The RFEA measurements are grey, the virtual IED sensor data are red.

*$N_2$.* The IEDs in N$_2$-plasma at a higher pressure (100 mTorr), measured with the RFEA and estimated with the virtual IED sensor, are presented in figure 8. Since the sheath in N$_2$-plasma is thicker than in Ar, the IED shape is more "collisional", i.e. the rf peaks are less pronounced, and the fraction of low-energy ions is much higher. As can be seen, the results are in good agreement, with the exception of the low-energy region of ~ 0 ÷ 20 eV, but, as already noted, the RFEA may not operate correctly in this energy range.

As a small conclusion to this subsection, we can state that even in the presence of collisions in the sheath, the virtual IED sensor works quite well. Moreover, the accuracy of these estimations is practically not inferior to the case of a slightly collisional sheath observed at a pressure of 20 mTorr.

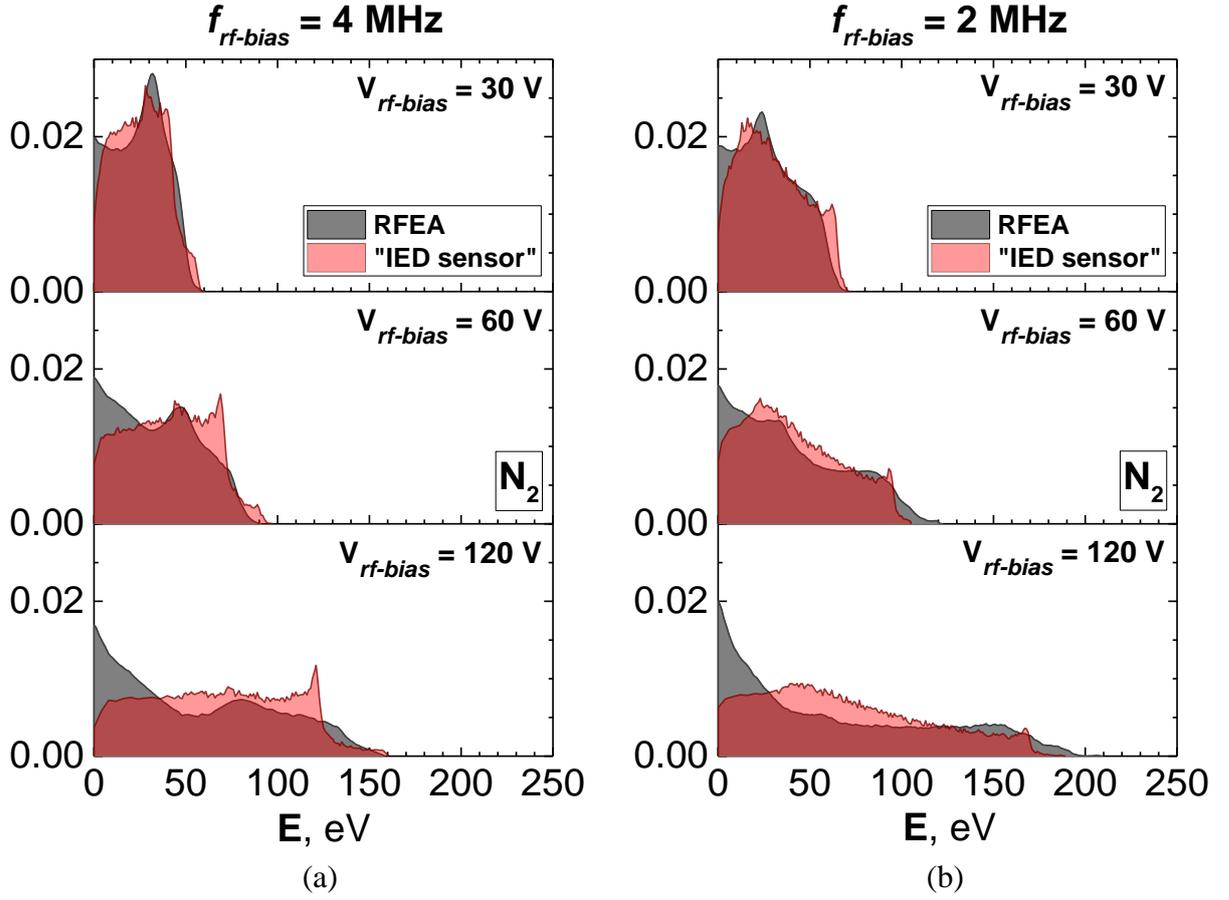

**Figure 8.** The virtual IED sensor validation in $N_2$-plasma ($n_e = 3 \cdot 10^9$ cm$^{-3}$) at a pressure of 100 mTorr and different rf-bias parameters: (a) $f_{rf-bias} = 4$ MHz; (b) $f_{rf-bias} = 2$ MHz. The RFEA measurements are grey, the virtual IED sensor data are red.

### 4.3 Effect of sheath thickness

Choosing the exact value for the sheath thickness $s_m$ is always a challenge when it comes to calculating the IED. It is often used as a variable parameter or estimated according to the well-known Child-Langmuir law (3.3). The concept of the virtual IED sensor assumes that $s_m$ is an input parameter, so there should be a method to accurately estimate $s_m$ using measurable discharge parameters. The purpose of this section is to find the most appropriate one.

It is worth noting that even in the collisionless case, the sheath thickness $s_m$ depends on the sheath voltage waveform $V_s(t)$. This is a consequence of the fact that $V_s(t)$ determines that part of the rf period, during which electrons fill the layer of uncompensated ion charge $q_i$. The sheath thickness $s_m$ is determined by the value of this $q_i$ (for a given $V_s(t)$). From an electrical point of view, this is equivalent to the simplest condition: $q_i/C_s = V_s = const$, where the sheath capacitance $C_s$ at fixed plasma parameters (gas pressure, plasma density, etc.) is inversely proportional to the sheath thickness $s_m$. Taking these considerations into account, let us perform a qualitative analysis of $s_m$ depending on the sheath voltage waveform $V_s(t)$ (the peak-to-peak sheath voltage amplitude $V_{s-pp}$ is fixed) in the following cases:

- $V_s(t) = const$, i. e. the case of a dc sheath;
- $V_s(t)$ is sinusoidal, i. e. the case of an rf sheath in a symmetric discharge;
- $V_s(t)$ has a "peaks" or "valleys" waveform, i. e. the case of an rf sheath in an asymmetric discharge, as well as in a symmetric discharge, where the rf-bias voltage waveform tailoring is specially used to control the IED [37,38,40,51]. It is important to notice that in [37,38,51] the voltage waveforms were determined relative to a grounded electrode. If an rf-biased electrode is considered as in this paper (figure 1), the terms "peaks" and "valleys" must be reversed. Then the $V_s(t)$ waveform in figure 2(b) corresponds to the "peaks" waveform. This terminology will be used below.

In the case of a dc sheath, the sheath thickness $s_m$ will be the minimum possible, since the uncompensated ion charge $q_i$ is maximum. $s_m$ can be determined by the Child-Langmuir law for a collisionless dc sheath (3.3) with $K_i = 4/9$.

In the case of a symmetric rf discharge with a sinusoidal sheath voltage waveform $V_s(t)$, the layer of uncompensated ion charge $q_i$ is periodically filled with electrons, so $q_i$ will be less, and the sheath thickness $s_m$ will be greater than in the "dc sheath" case. $s_m$ should be estimated using the Child-Langmuir law (3.3) with $K_i = 0.82$, which corresponds to a collisionless rf sheath.

In the case of a "peaks" $V_s(t)$ waveform (see figure 2(b)), electrons fill the layer of uncompensated ion charge $q_i$ for a smaller part of the rf period than in the case of a sinusoidal $V_s(t)$. Thus, $s_m$ will be greater than in the "dc sheath" case, but less than in the "rf sheath" case. But if $V_s(t)$ represents a "valleys" waveform, then electrons fill the layer for most of the rf period compared to the sinusoidal $V_s(t)$ case, which results in a larger $s_m$ value.

To summarize: the uncompensated ion charge $q_i^{dc} > q_i^{"peaks"} > q_i^{rf} > q_i^{"valleys"}$ and the sheath thickness $s_m^{dc} < s_m^{"peaks"} < s_m^{rf} < s_m^{"valleys"}$.

In this work, the following methods for estimating $s_m$ are used:

- the Child-Langmuir law for a collisionless dc sheath (3.3) with $K_i = 4/9$ ($V_s(t) = \overline{V}_s = V_{dc} + \overline{V}_p$ [23], where $V_{dc}$ is the dc self-bias potential, $\overline{V}_p$ is the average plasma potential);
- the Child-Langmuir law for a collisionless rf sheath (3.3) with $K_i = 0.82$ ($V_s(t)$ is assumed to be sinusoidal, and $\overline{V}_s$ is calculated in the same way);
- varying $s_m$ as a parameter until the calculated IED (primarily its width) coincides with the IED measured under the same discharge conditions.

The case of zero applied rf-bias ($V_{rf-bias} = 0$ V) in the df rf CCP discharge can be considered as close to the "dc sheath" case, since the amplitude of rf 81 MHz oscillations in the sheath above the bottom electrode is small ($\sim \bar{V}_p$). Therefore, the Child-Langmuir law variations for the dc and rf sheaths are expected to form, respectively, the lower and upper boundaries of the range of possible $s_m$ values under the conditions of this experiment (no "valleys" $V_s(t)$ waveform was observed).

The results of estimating $s_m$ by these three methods are shown in figures 9-10(a, b) for Ar- and N$_2$-plasma at pressures of 20 and 100 mTorr and in figure 11 for Xe-plasma at 20 mTorr. In general, most of the $s_m$ values, obtained by fitting the calculated IEDs to the experimental ones, fall within the range formed by the Child-Langmuir formulae for the dc and rf sheaths. It can also be seen that this entire range shifts upward towards greater $s_m$ values at 100 mTorr (see figures 9(b) and 10(b)). At higher pressure, the ion flux $F_i$ to the electrode decreases due to collisions in the presheath region [15,19,52], which, in turn, leads to an increase in the sheath thickness $s_m$.

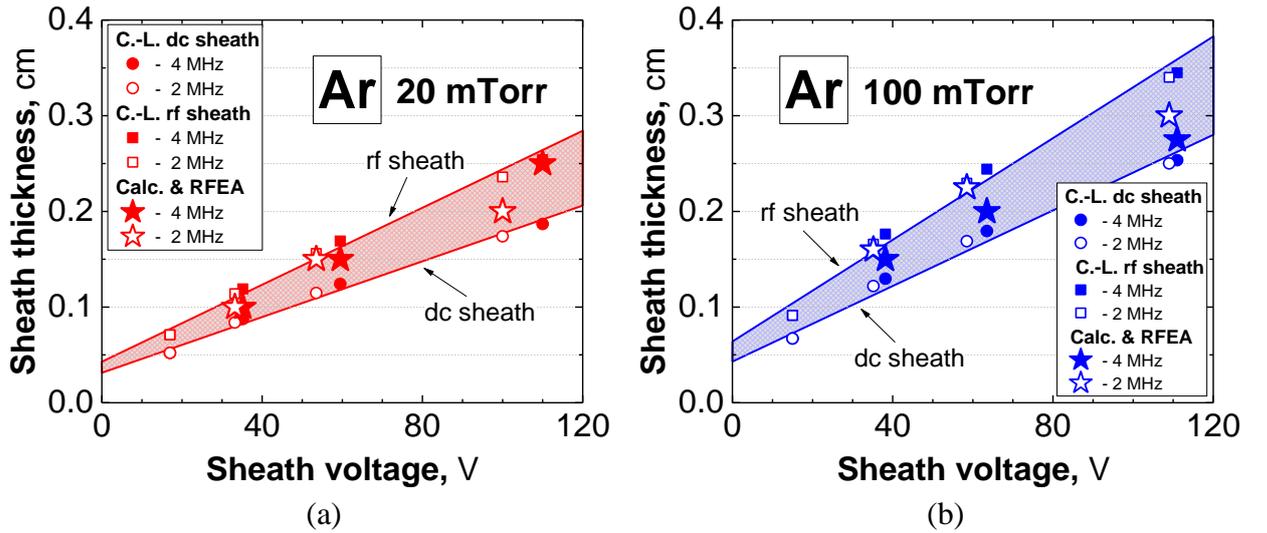

**Figure 9.** Sheath thickness $s_m$ dependence on the average sheath voltage $V_s$ in Ar-plasma ($n_e = 2 \cdot 10^{10}$ cm$^{-3}$) at a pressure of a) 20 mTorr, b) 100 mTorr. The circles show the $s_m$ estimations by the Child-Langmuir law for a dc sheath, and the squares correspond to the Child-Langmuir law for an rf sheath. The $s_m$ values, obtained by fitting the calculated IEDs to the measured ones, are marked with asterisks.

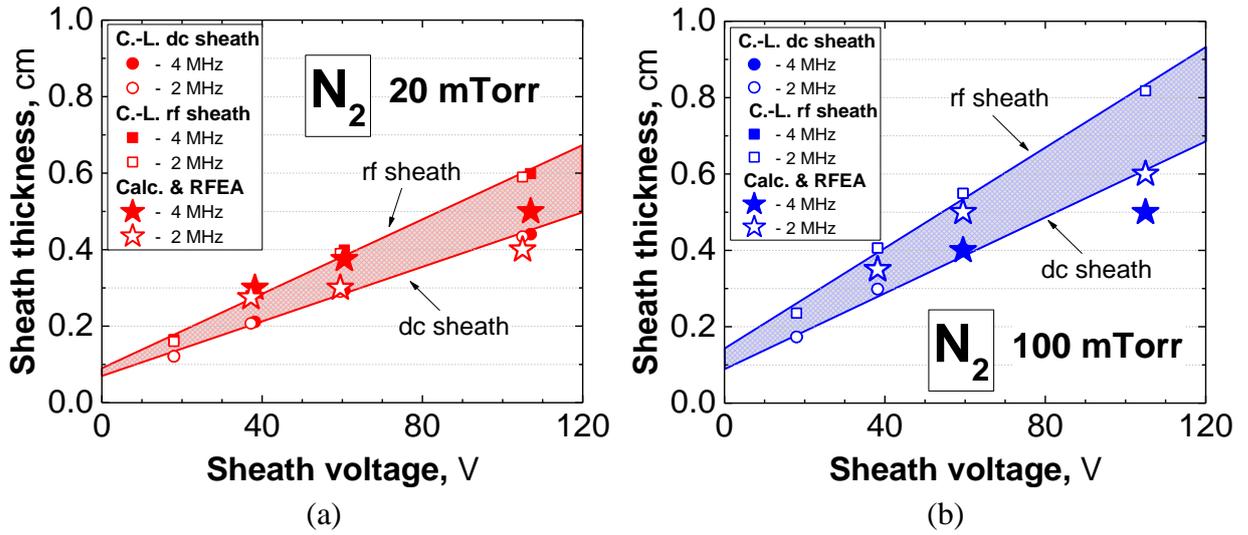

**Figure 10.** Sheath thickness $s_m$ dependence on the average sheath voltage $V_s$ in N$_2$-plasma ($n_e = 3 \cdot 10^9$ cm$^{-3}$) at a pressure of a) 20 mTorr, b) 100 mTorr. The circles show the $s_m$ estimations by the Child-Langmuir law for a dc sheath, and the squares correspond to the Child-Langmuir law for an rf sheath. The $s_m$ values, obtained by fitting the calculated IEDs to the measured ones, are marked with asterisks.

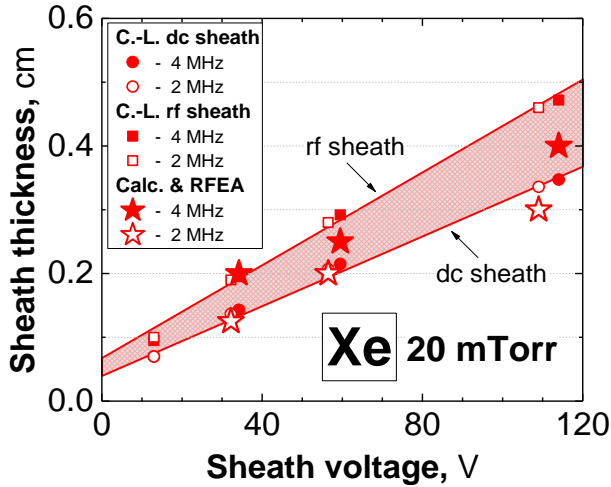

**Figure 11.** Sheath thickness $s_m$ dependence on the average sheath voltage $V_s$ in Xe-plasma ($n_e = 2 \cdot 10^{10}$ cm$^{-3}$) at a pressure of 20 mTorr. The circles show the $s_m$ estimations by the Child-Langmuir law for a dc sheath, and the squares correspond to the Child-Langmuir law for an rf sheath. The $s_m$ values, obtained by fitting the calculated IEDs to the measured ones, are marked with asterisks.

The next step is to check how sensitive the shape of the calculated IED is to the $s_m$ estimation method. The IED in Ar-plasma at a pressure of 20 mTorr and rf-bias parameters of 2 MHz and 60 V was calculated using the $s_m$ values obtained by three methods. The results are presented in figure 12 along with the IED measured by the RFEA under the same conditions. It is seen that the smaller the $s_m$ value, the wider the IED: the one that is calculated using $s_m$, estimated by the Child-Langmuir law for the dc sheath case, is the widest. On the whole, these changes in the IED width are small.

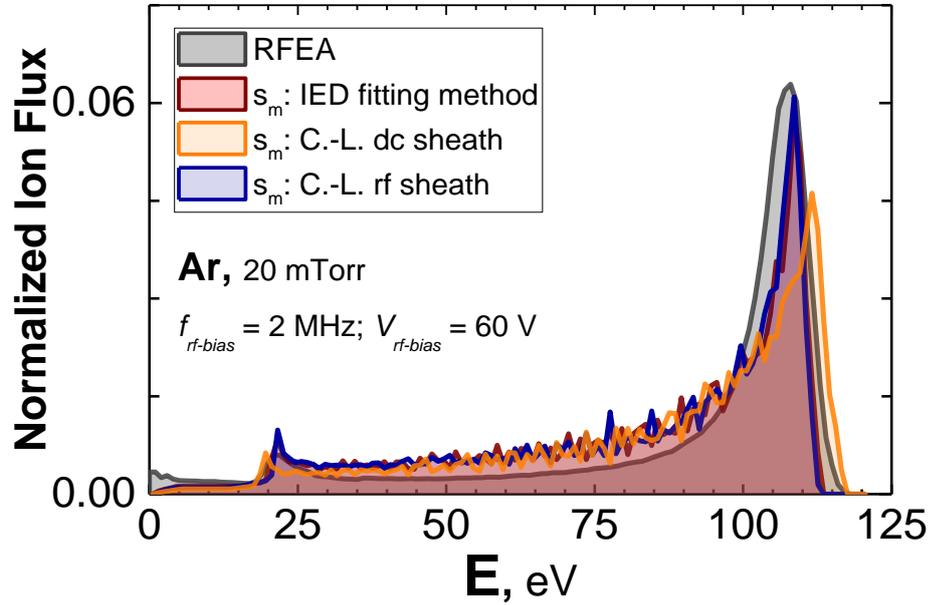

**Figure 12.** IEDs in Ar-plasma ($n_e = 2 \cdot 10^{10}$ cm$^{-3}$), obtained with the "virtual IED sensor", depending on estimations of the sheath thickness $s_m$ in three ways: IED fitting method – dark red line; estimations by the Child-Langmuir law for a dc sheath – orange line; estimations by the Child-Langmuir law for an rf sheath – blue line. Direct measurements with the RFEA are also shown for comparison (dark grey line).

Based on the results demonstrated in figures 9-12, it can be concluded that the Child-Langmuir law gives an appropriate estimation for the $s_m$ value, which can be successfully used for the virtual IED sensor. The accuracy of the sensor may be slightly increased by choosing the coefficient in the Child-Langmuir formula (3.3) in accordance with the $V_s(t)$ waveform: $K_i = 4/9$, if it is closer to a constant, or $K_i = 0.82$, if it is closer to the sine. But what is much more important for a correct $s_m$ estimation is to accurately measure the ion flux $F_i$ or the plasma density $n_e$ [15].

### 4.4 Effect of sheath voltage waveform

Although the sheath voltage waveform $V_s(t)$ is not so important for estimating the sheath thickness $s_m$, it still strongly affects the IED shape. The sheath voltage waveform $V_s(t)$ is an input parameter along with the sheath thickness $s_m$.

The virtual IED sensor results in sections 4.1 and 4.2 show that the IED shape is reproduced quite accurately when the actual sheath voltage waveform $V_s(t)$ (distorted relative to the sine due to asymmetric discharge geometry) is taken into account. For this purpose, the potential waveforms of both the electrode $V_{el}(t)$ and the plasma $V_p(t)$ are to be measured (see section 2.2). But it is clear that the best diagnostics is the one that requires fewer additional measurements. Not to mention, plasma processing reactors do not offer the possibility of a large number of measurements, unlike research discharge chambers. For this reason, it is useful to check whether the sheath voltage waveform $V_s(t)$ is indeed strongly reflected in the IED shape or it can simply be replaced with a sinusoidal waveform.

There is a third compromise option to take into account the sheath voltage waveform $V_s(t)$. This option avoids measuring the plasma potential waveform $V_p(t)$, but still takes into account the effect of the sheath voltage waveform $V_s(t)$ distortion on the IED shape. The electrode potential waveform $V_{el}(t)$ is easy to measure because it does not require the introduction of an rf antenna into the plasma bulk, which is necessary to obtain the plasma potential waveform $V_p(t)$. Thus, the idea is to use formula (2.1), but instead of the measured plasma potential waveform $V_p(t)$, use the calculated one:

$$V_p(t) = \overline{V_p}(1 + \sin(2\pi f_h t)) , \qquad (4.1)$$

where $f_h$ is the high frequency or plasma generation frequency of 81 MHz, and the average value of the plasma potential $\overline{V_p}$ is estimated, for example, from the assumed electron temperature $T_e$. The point is that $T_e$ in most low-pressure plasmas used for processing varies within a limited range, usually $\sim 2 \div 6$ eV. So, the average plasma potential $\overline{V_p}$ value is also in a limited range of $\sim 20 \div 35$ V, and the influence of the inaccuracy of the $\overline{V_p}$ estimation on the IED position on the energy scale will be insignificant.

The effect of the sheath voltage waveform $V_s(t)$ on the IED calculation was observed in Ar-plasma at 20 mTorr and rf-bias parameters of 2 MHz and 60 V. Three different $V_s(t)$ waveforms were taken: measured, mixed and calculated as a sine. Each $V_s(t)$ was calculated by (2.1) using the corresponding electrode $V_{el}(t)$ and plasma $V_p(t)$ potential waveforms, which are presented in figure 13 (from top to bottom):

- both $V_{el}(t)$ and $V_p(t)$ are measured;
- $V_{el}(t)$ is measured and $V_p(t)$ is calculated as a sine by (4.1);
- both $V_{el}(t)$ and $V_p(t)$ are calculated. $V_{el}(t)$ is calculated as follows:

$$V_{el}(t) = V_{dc} + V_{rf-bias}\sin(2\pi f_{rf-bias} t), \qquad (4.2)$$

where $V_{dc}$ is the dc self-bias voltage that is measured outside the chamber (the measurement point is marked in figure 1), and $V_{rf-bias}$ and $f_{rf-bias}$ are the known voltage and frequency of the applied rf-bias, respectively.

The $V_{el}(t)$ and $V_p(t)$ waveforms are calculated under the assumption that there is a frequency decoupling, i.e. there is no rf-bias frequency modulation of the plasma potential $V_p(t)$, as well as there is no high-frequency modulation of the electrode potential $V_{el}(t)$. Although there is a slight low-frequency modulation of the measured $V_p(t)$, which can be seen in figure 13 (top of the figure), it will certainly not have a visible effect on the IED calculation. The mixed $V_s(t)$ waveform was obtained using the low-frequency component $V_{el}^{2\,MHz}(t)$ of the measured $V_{el}(t)$ waveform, following the same assumption.

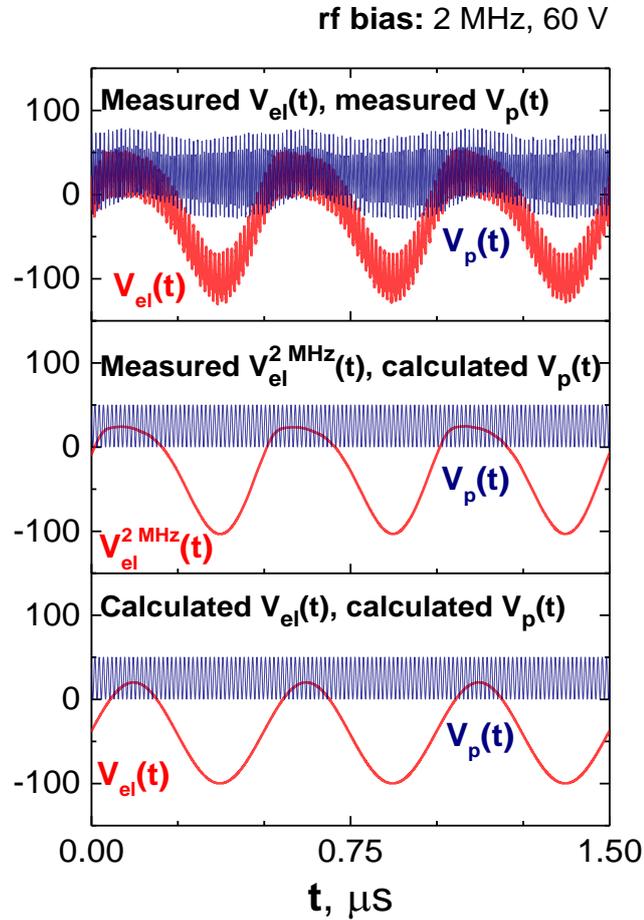

**Figure 13.** The electrode $V_{el}(t)$ and plasma $V_p(t)$ potential waveforms, of which three different $V_s(t)$ waveforms – from top to bottom: measured, mixed, and calculated – were obtained using (2.1). All of the potential waveforms are presented for Ar-plasma at 20 mTorr and rf-bias parameters of 2 MHz and 60 V.

The results of the IED calculation with these three $V_s(t)$ waveforms – measured, mixed, and calculated as a sine – are demonstrated in figure 14 along with the IED, measured by the RFEA. As expected, the best agreement with experiment (grey line) is observed when the IED is calculated using the measured $V_s(t)$ waveform (dark red line), and the calculation with the sinusoidal $V_s(t)$ waveform (blue line) gives the largest discrepancy with experimental data. The compromise option of the mixed $V_s(t)$ waveform gives something in between: it still takes into account the effect of the non-sinusoidal $V_s(t)$ waveform, but not completely.

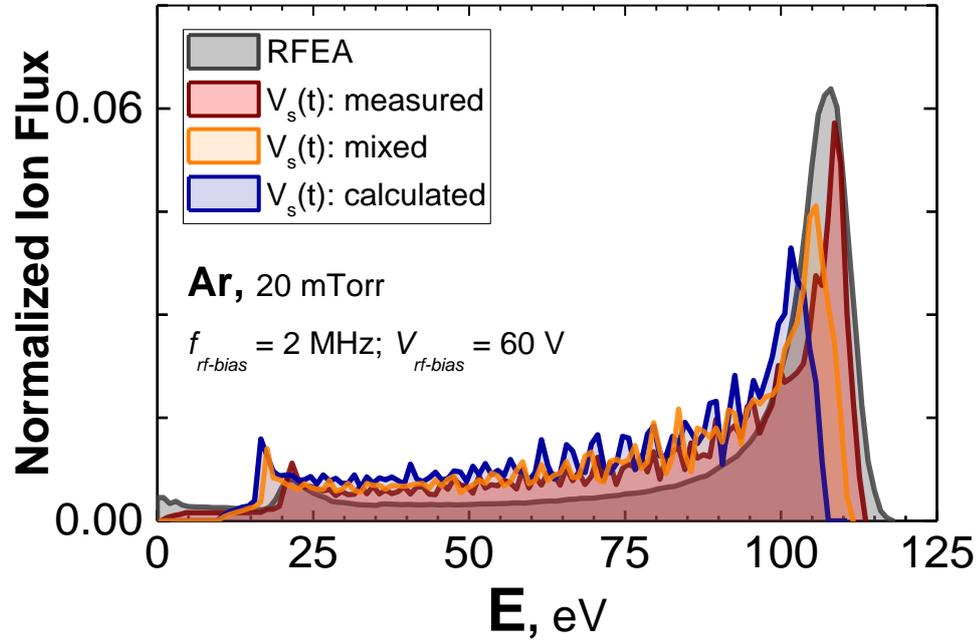

**Figure 14.** IEDs in Ar-plasma ($n_e = 2 \cdot 10^{10}$ cm$^{-3}$), obtained using the "virtual IED sensor" depending on the sheath voltage waveform $V_s(t)$: measured – dark red line; mixed, i. e. the electrode potential waveform $V_{el}(t)$ is measured, and the plasma potential waveform $V_p(t)$ is calculated as a sine by (4.1) – orange line; calculated as a sine – blue line. Direct measurements via the RFEA are also shown for comparison (dark grey line).

The conclusion to this section is quite simple: the more information on the actual sheath voltage waveform $V_s(t)$, the more accurately the IED shape is reproduced. If it is not possible to measure both the plasma $V_p(t)$ and the electrode $V_{el}(t)$ potential waveforms, then even measuring $V_{el}(t)$ only and estimating $V_p(t)$ from the discharge conditions will result in a more accurate IED estimation.

### 4.5 "Real-time" operating regime

One of the key characteristics of the virtual IED sensor is its speed, or the average time it takes to estimate one IED. Typical plasma processing times are > 1 min to provide the best process control. So, the estimation time should be short enough to ensure its continuous operation in real time ($\sim 1 \div 10$ s). At the same time, the data quality should not fall due to the high speed of the sensor.

Figure 15 shows three results of calculating an IED with different numbers of particles used in the fast IED calculation method. Calculation of $10^5$ particles provides a smoother IED, but takes more time (~ 1 min for 1 IED), while calculation of $10^4$ particles gives an IED of almost the same shape, although a little noisier. The calculation time in the second case is ~ 10 sec.

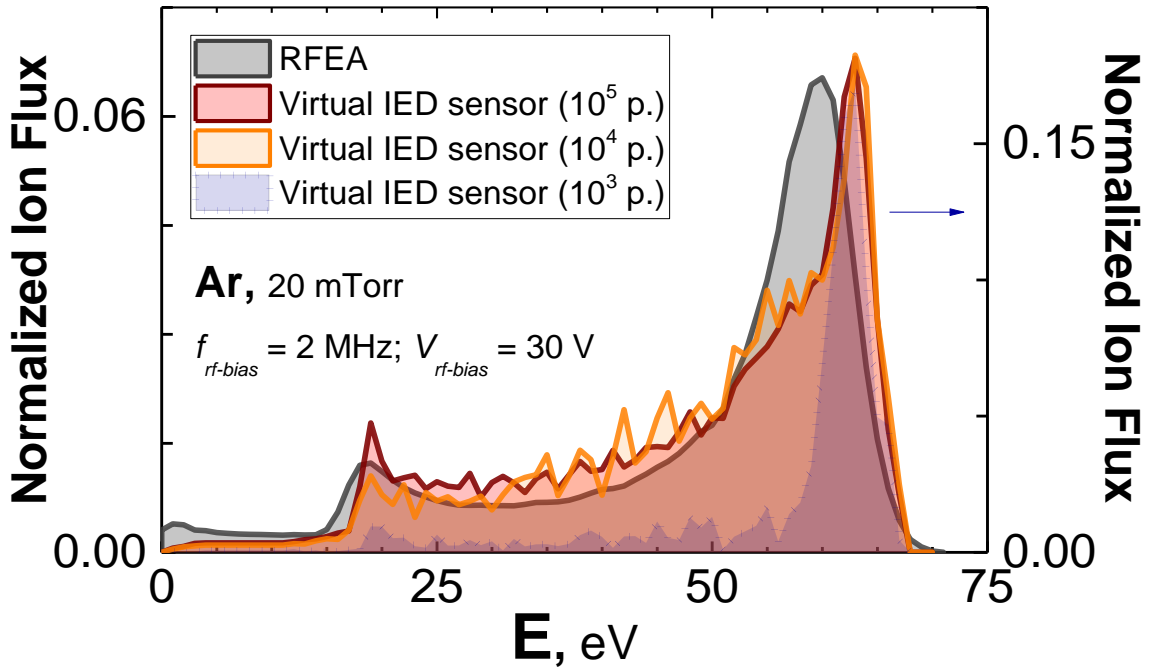

**Figure 15.** IEDs in Ar-plasma ($n_e = 2 \cdot 10^{10}$ cm$^{-3}$) obtained via the virtual IED sensor depending on the calculation statistics ($10^5$ vs $10^4$ vs $10^3$ particles). Direct measurements via the RFEA are also shown for comparison.

The shortest time among the cases considered is required to calculate $10^3$ particles, but the IED shape is very different from that of other IEDs - measured with the RFEA and calculated with a larger number of particles ($\geq 10^4$). The main peaks arising from the ion response to the rf-bias field are reproduced, but most of the lower-energy ions are not taken into account. The fact is that the total number of $10^3$ particles reaching the electrode is provided by more energetic ions. In other words, with such a total number of particles, the low-energy ions do not have enough time to reach the electrode. So, the proportion between the numbers of low-energy and high-energy ions is less than in the case of higher particle statistics ($\geq 10^4$).

To summarize, we can say that the most adequate number of particles to use in calculations in terms of the IED estimation time and quality is $10^4$: the IED is calculated with satisfactory accuracy in ~ 10 sec. Of course, the concept of "real time" implies a much higher sensor speed. But here the "real time" operating regime of the virtual IED sensor means rather continuous updating of the IED information depending on the measured discharge parameters. The results of this section showed that the virtual IED sensor is quite capable of this task.

## 5 Conclusion

Direct diagnostics of the ion energy distribution at the surface of an rf-biased electrode during material plasma processing is difficult for a number of reasons. That is why, the concept of the "virtual IED sensor", i. e. fast "real-time" IED calculations using constantly updated information on the plasma parameters and discharge conditions, can be considered as an alternative.

In this paper, the "virtual IED sensor" has been validated in an asymmetric df rf CCP discharge in Ar, $N_2$ and Xe. The results showed that the idea works: the sensor allows the IED trends to be monitored depending on changing discharge conditions such as gas sort and pressure, plasma density, rf-bias frequency and voltage amplitude. The fast IED calculation method underlying the sensor takes into account the effect of discharge asymmetry and ion-neutral collisions in the sheath on the IED shape. However, the accuracy of the IED estimations by the sensor is directly dependent on the amount and quality of information about the plasma process used as input data.

Generally speaking, the question of the accuracy of the IED estimations is relative. It should be discussed primarily in terms of quantities that are important for a particular application. The error in the IED estimation can be judged in different ways: in terms of visual inaccuracy of the calculated IED compared to the directly measured by the RFEA, or more specific inaccuracies – in the average or maximum ion energy or in the energy of the IED peaks. In some applications, for example, the maximum ion energy in the spectra is important for initiating threshold processes, while ions of too high energies are undesirable, since they can damage the treated surface.

In general, it should be understood that most plasma treatment processes require not so much an accurate reproduction of the IED shape, but the ability to track how it changes with discharge conditions and plasma parameters. And for this purpose it is quite possible to use the presented virtual IED sensor.

## Acknowledgements

The reported study was funded by RFBR according to the research project № 18-29-27003 and the Interdisciplinary Scientific and Educational School of Moscow University "Photonic and Quantum Technologies. Digital Medicine".

# References

[1]     Goto H H, Löwe H and Ohmi T 1992 Dual excitation reactive ion etcher for low energy plasma processing *J. Vac. Sci. Technol. A* **10** 3048–54

[2]     Goto H H, Löwe H and Ohmi T 1993 Independent control of ion density and ion bombardment energy in a dual RF excitation plasma *IEEE Trans. Semicond. Manuf.* **6** 58–64

[3]     Bi Z, Liu Y, Jiang W, Xu X and Wang Y 2011 A brief review of dual-frequency capacitively coupled discharges *Curr. Appl. Phys.* **11** S2–8

[4]     Rakhimova T V, Braginsky O V, Ivanov V V, Kim T K, Kong J T, Kovalev A S, Lopaev D V, Mankelevich Y A, Proshina O V and Vasilieva A N 2006 Experimental and theoretical study of RF plasma at low and high frequency *IEEE Trans. Plasma Sci.* **34** 867–77

[5]     Wu A C F, Lieberman M A and Verboncoeur J P 2007 A method for computing ion energy distributions for multifrequency capacitive discharges *J. Appl. Phys.* **101** 056105

[6]     Sharma S, Mishra S K, Kaw P K and Turner M M 2017 The effect of intermediate frequency on sheath dynamics in collisionless current driven triple frequency capacitive plasmas *Phys. Plasmas* **24** 013509

[7]     Heil B G, Czarnetzki U, Brinkmann R P and Mussenbrock T 2008 On the possibility of making a geometrically symmetric RF-CCP discharge electrically asymmetric *J. Phys. D. Appl. Phys.* **41** 165202

[8]     Lafleur T 2015 Tailored-waveform excitation of capacitively coupled plasmas and the electrical asymmetry effect *Plasma Sources Sci. Technol.* **25** 13001

[9]     Economou D J 2014 Pulsed plasma etching for semiconductor manufacturing *J. Phys. D. Appl. Phys.* **47** 303001

[10]    Chen Z, Longo R C, Hummel M, Carruth M, Blakeney J, Ventzek P, Ranjan A and Ranjan A 2020 Factors influencing ion energy distributions in pulsed inductively coupled argon plasmas *J. Phys. D. Appl. Phys.* **53** 335202

[11]    Lee J B, Chang H Y and Seo S H 2013 Collisional effect on the time evolution of ion energy distributions outside the sheath during the afterglow of pulsed inductively coupled plasmas *Plasma Sources Sci. Technol.* **22** 065008

[12]    Voloshin D, Kovalev A, Mankelevich Y, Proshina O, Rakhimova T and Vasilieva A 2015 Evaluation of plasma density in RF CCP discharges from ion current to Langmuir probe: Experiment and numerical simulation *Eur. Phys. J. D* **69** 23

[13]    Voloshin D, Rakhimova T and Mankelevich Y 2016 The plasma sheath around planar probes: Effects of ion collisions *Plasma Sources Sci. Technol.* **25** 015018

[14]   Voloshin D G, Vasil'eva A N, Kovalev A S, Mankelevich Y A and Rakhimova T V 2016 Determination of plasma density from data on the ion current to cylindrical and planar probes *Plasma Phys. Reports* **42** 1146–54

[15]   Bogdanova M, Lopaev D, Zyryanov S, Voloshin D and Rakhimova T 2018 Relation between the ion flux and plasma density in an rf CCP discharge *Plasma Sources Sci. Technol.* **27** 025003

[16]   Bogdanova M, Lopaev D, Zyryanov S, Voloshin D and Rakhimova T 2019 Ion composition of rf CCP in Ar/H2 mixtures *Plasma Sources Sci. Technol.* **28** 095017

[17]   Bogdanova M A, Lopaev D V, Zyryanov S M and Rakhimov A T 2016 "Virtual IED sensor" at an rf-biased electrode in low-pressure plasma *Phys. Plasmas* **23** 073510

[18]   Kawamura E, Vahedi V, Lieberman M A and Birdsall C K 1999 Ion energy distributions in rf sheaths; review, analysis and simulation *Plasma Sources Sci. Technol.* **8** R45

[19]   Lieberman M A and Lichtenberg A J 2005 *Principles of Plasma Discharges and Materials Processing* (Hoboken: John Wiley & Sons, Inc.)

[20]   Benoit-Cattin P and Bernard L 1968 Anomalies of the Energy of Positive Ions Extracted from High-Frequency Ion Sources. A Theoretical Study *J. Appl. Phys.* **39** 5723–6

[21]   Sobolewski M A, Wang Y and Goyette A 2002 Measurements and modeling of ion energy distributions in high-density, radio-frequency biased CF4 discharges *J. Appl. Phys.* **91** 6303

[22]   Hayden C, Gahan D and Hopkins M B 2009 Ion energy distributions at a capacitively and directly coupled electrode immersed in a plasma generated by a remote source *Plasma Sources Sci. Technol.* **18** 25018

[23]   Gahan D, Daniels S, Hayden C, O'Sullivan D O and Hopkins M B 2012 Characterization of an asymmetric parallel plate radio-frequency discharge using a retarding field energy analyzer *Plasma Sources Sci. Technol.* **21** 015002

[24]   Chen W C and Pu Y K 2014 An analytical model for time-averaged ion energy distributions in collisional rf sheaths *J. Phys. D. Appl. Phys.* **47** 345201

[25]   Israel D, Riemann K U and Tsendin L 2006 Charge exchange collisions and ion velocity distribution at the electrode of low pressure capacitive rf discharges *J. Appl. Phys.* **99** 093303

[26]   Diomede P, Economou D J and Donnelly V M 2012 Rapid calculation of the ion energy distribution on a plasma electrode *J. Appl. Phys.* **111** 123306

[27]   Charles C, Degeling A W, Sheridan T E, Harris J H, Lieberman M A and Boswell R W 2000 Absolute measurements and modeling of radio frequency electric fields using a retarding field energy analyzer *Phys. Plasmas* **7** 5232–41

[28]   Donkó Z, Schulze J, Czarnetzki U, Korolov I, Hartmann P, Derzsi A and Schüngel E 2012 Fundamental investigations of capacitive radio frequency plasmas: simulations and experiments *Plasma Phys. Control. Fusion* **54** 124003

[29]   Faraz T, Arts K, Karwal S, Knoops H C M and Kessels W M M 2019 Energetic ions during plasma-enhanced atomic layer deposition and their role in tailoring material properties *Plasma Sources Sci. Technol.* **28** 24002

[30]   Chabert P and Braithwaite N 2011 *Physics of radio-frequency plasmas* (Cambridge: Cambridge University Press)

[31]   Ellmer K, Wendt R and Wiesemann K 2003 Interpretation of ion distribution functions measured by a combined energy and mass analyzer *Int. J. Mass Spectrom.* **223**–**224** 679–93

[32]   Baloniak T, Reuter R, Flötgen C and Von Keudell A 2010 Calibration of a miniaturized retarding field analyzer for low-temperature plasmas: Geometrical transparency and collisional effects *J. Phys. D. Appl. Phys.* **43** 055203

[33]   Talley M L, Shannon S, Chen L and Verboncoeur J P 2017 IEDF distortion and resolution considerations for RFEA operation at high voltages *Plasma Sources Sci. Technol.* **26** 125001

[34]   van de Ven T H M, de Meijere C A, van der Horst R M, van Kampen M, Banine V Y and Beckers J 2018 Analysis of retarding field energy analyzer transmission by simulation of ion trajectories *Rev. Sci. Instrum.* **89** 43501

[35]   Schulze J, Schüngel E, Donkó Z and Czarnetzki U 2011 The electrical asymmetry effect in multi-frequency capacitively coupled radio frequency discharges *Plasma Sources Sci. Technol.* **20** 15017

[36]   Lafleur T and Booth J P 2012 Control of the ion flux and ion energy in CCP discharges using non-sinusoidal voltage waveforms *J. Phys. D. Appl. Phys.* **45** 395203

[37]   Lafleur T, Delattre P A, Johnson E V and Booth J P 2012 Separate control of the ion flux and ion energy in capacitively coupled radio-frequency discharges using voltage waveform tailoring *Appl. Phys. Lett.* **101** 124104

[38]   Delattre P A, Lafleur T, Johnson E and Booth J P 2013 Radio-frequency capacitively coupled plasmas excited by tailored voltage waveforms: Comparison of experiment and particle-in-cell simulations *J. Phys. D. Appl. Phys.* **46** 235201

[39]   Lafleur T, Delattre P A, Johnson E V and Booth J P 2013 Capacitively coupled radio-frequency plasmas excited by tailored voltage waveforms *Plasma Phys. Control. Fusion* **55** 124002

[40]   Faraz T, Verstappen Y G P, Verheijen M A, Chittock N J, Lopez J E, Heijdra E, Van

Gennip W J H, Kessels W M M and MacKus A J M 2020 Precise ion energy control with tailored waveform biasing for atomic scale processing *J. Appl. Phys.* **128** 213301

[41]  Donkó Z, Schulze J, Hartmann P, Korolov I, Czarnetzki U and Schüngel E 2010 The effect of secondary electrons on the separate control of ion energy and flux in dual-frequency capacitively coupled radio frequency discharges *Appl. Phys. Lett.* **97** 081501

[42]  Gans T, Schulze J, O'Connell D, Czarnetzki U, Faulkner R, Ellingboe A R and Turner M M 2006 Frequency coupling in dual frequency capacitively coupled radio-frequency plasmas *Appl. Phys. Lett.* **89** 261502

[43]  Biondi M A 1951 Measurement of the Electron Density in Ionized Gases by Microwave Techniques *Rev. Sci. Instrum.* **22** 500–2

[44]  Stenzel R L 1976 Microwave resonator probe for localized density measurements in weakly magnetized plasmas *Rev. Sci. Instrum.* **47** 603–7

[45]  Piejak R B, Godyak V A, Garner R, Alexandrovich B M and Sternberg N 2004 The hairpin resonator: A plasma density measuring technique revisited *J. Appl. Phys.* **95** 3785–91

[46]  Schüngel E, Donkó Z and Schulze J 2017 A Simple Model for Ion Flux-Energy Distribution Functions in Capacitively Coupled Radio-Frequency Plasmas Driven by Arbitrary Voltage Waveforms *Plasma Process. Polym.* **14** 1600117

[47]  Skullerud H R 1968 The stochastic computer simulation of ion motion in a gas subjected to a constant electric field *J. Phys. D. Appl. Phys.* **1** 1567–8

[48]  Phelps A V. 1994 The application of scattering cross sections to ion flux models in discharge sheaths *J. Appl. Phys.* **76** 747–53

[49]  Phelps A V. 1991 Cross Sections and Swarm Coefficients for Nitrogen Ions and Neutrals in N2 and Argon Ions and Neutrals in Ar for Energies from 0.1 eV to 10 keV *J. Phys. Chem. Ref. Data* **20** 557–73

[50]  Babaeva N Y, Lee J K and Shon J W 2005 Capacitively coupled plasma source operating in Xe/Ar mixtures *J. Phys. D. Appl. Phys.* **38** 287–99

[51]  Diomede P, Economou D J, Lafleur T, Booth J P and Longo S 2014 Radio-frequency capacitively coupled plasmas in hydrogen excited by tailored voltage waveforms: Comparison of simulations with experiments *Plasma Sources Sci. Technol.* **23** 065049

[52]  Lafleur T and Chabert P 2015 Edge-to-center density ratios in low-temperature plasmas *Plasma Sources Sci. Technol.* **24** 025017